# Low-Cost Performance-Efficient Field-Programmable Pin-Constrained Digital Microfluidic Biochip

Alireza Abdoli, Sedigheh Farhadtoosky and Ali Jahanian*

Department of Computer Science and Engineering

Shahid Beheshti University, G.C., Tehran, Iran

{al.abdoli, s.farhadtoosky}@mail.sbu.ac.ir, jahanian@sbu.ac.ir

## ABSTRACT

Digital microfluidic biochips (DMFBs) are revolutionary biomedical devices towards diagnostics and point-of-care applications; the chips provide the capability of performing wide ranges of biochemistry and laboratory procedures, offering various opportunities among which to mention are automation, miniaturization and cost-affordability of bioassays. There have been various digital microfluidic biochips architectures; the application-specific chips are mainly suited towards executing a predefined set of bioassays whereas the more flexible general-purpose chips allow executing wide ranges of bioassays on the same architecture. Though more flexible in terms of performing various bioassays the general-purpose chips require more complicated designs compared with application-specific counterparts necessitating larger and more costly designs. This paper attempts to propose a general-purpose field-programmable pin-constrained DMFB design with improved characteristics in terms area-consumption, manufacturing cost and performance.

## 1. INTRODUCTION

Digital microfluidic biochips (DMFBs) are modern generation of devices enabling a new paradigm in biochemistry/laboratory procedures and healthcare diagnostics. DMFBs offer various great opportunities never existed before all in one place; the most crucial advantages of DMFBs compared with prior methodologies include automation, miniaturization and cost-affordability. The chips are capable of performing wide range of fluidic bioassays among which are the following: polymerase chain reaction (PCR) [1], in-vitro diagnostics [2], protein-crystallization [3], and DNA computing [4].

DMFBs operate on the basis of manipulating fluid droplets with pico-liter to Nano-liter volumes on a 2D grid of electrodes. A typical DMFB consists of two (bottom and top) plates. The bottom plates consist of 2D array of electrodes on which the microfluidic operations are performed. The top plate comprises of single electrode spanning the bottom plate acting as the ground electrode. Both the top and bottom plates are coated with a dielectric layer (e.g. Parylene C) and a hydrophobic layer facilitating movement of droplets. Figure 1 (a) shows the cross-sectional view of a DMFB, further Figure 1 (b) illustrates the bottom plate of a DMFB with I/O reservoirs on the periphery of the array of electrodes. The droplets are sandwiched between the two plates. This is to facilitate movement of droplets the space between the top and bottom plates.

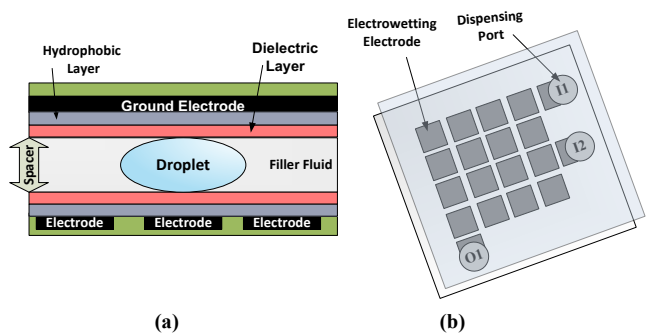

Figure 1 (a) Cross-sectional and (b) top view of a DMFB.

Over the course of past few years there have been numerous research articles on DMFB designs capable of general-purpose bioassay execution while at the same time lowering manufacturing costs. This study attempts to propose a low-cost general-purpose field-programmable DMFB design.

The rest of paper is organized as follows: Section 2 reviews the technology, synthesis flow and fundamental operations of DMFBs. Section 3 provides literature review on previous and state-of-art DMFB designs. Section 4 presents the proposed low-cost performance-efficient pin-constrained DMFB design. Section 5 provides analysis on hardware cost associated with manufacturing of DMFB designs. Finally, Section 6 concludes the paper.

## 2. BACKGROUND
### 2.1 Microfluidic Biochip Technology

There are various technologies for actuation of droplets on Digital (droplet-based) microfluidic biochips. One approach is through electrowetting on dielectric (EWOD) phenomenon [4]; which refers to electromechanical actuation (wetting) of conductive fluids on a solid surface through electrical bias [5]. Accordingly, droplets are actuated by applying appropriate voltage levels to the electrode beneath the droplet; which in creates an electrical field over the droplet. Subsequently, the tension between the droplet and surface of the electrode is increased which is used towards actuation of droplets.

Figure 1 illustrates the electrowetting on dielectric (EWOD) phenomenon. Recently, there have been alternative approaches to actuation of droplets in droplet-based microfluidic biochips. One such alternative approach is tilting DMFB [6]; which utilizes stepper motors towards mechanical agitation of droplets. Another alternative approach to EWoD-based DMFBs is through magnetic actuation of floating liquid marbles [7].

## 2.2 Fundamental DMFB Operations

Every bioassay consists of various microfluidic operations, such as temporary storing (holding) the droplet in place, Droplet transportation, mixing of different samples, dispense (sample introduction), merging and splitting. Many bioassays require thermal cycling, fluorescence detection and etc. These operations demand special equipment which must be accounted for during the design and manufacturing processes of the DMFB. An example would be an external thermal measurement module for microfluidic devices.

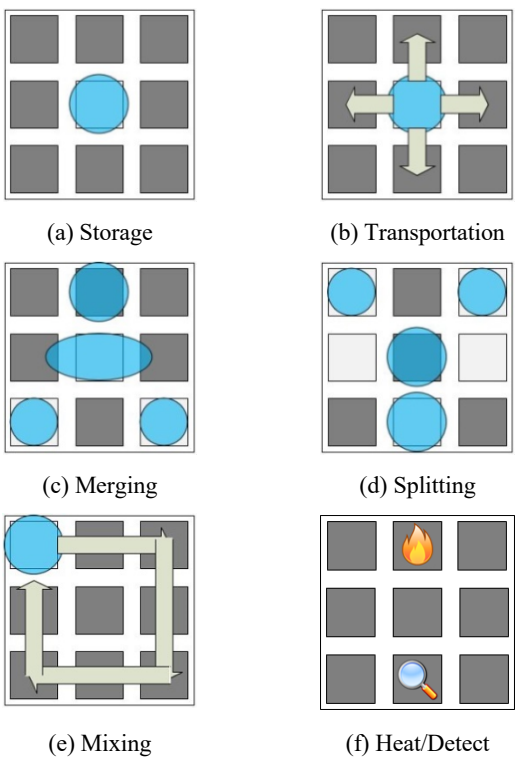

(a) Storage　　(b) Transportation
(c) Merging　　(d) Splitting
(e) Mixing　　(f) Heat/Detect

**Figure 2 Microfluidic operations**

## 2.3 Fluidic Level Synthesis

Accomplishment of bioassays on a typical DMFB involves various input and synthesis stages. The process is commenced with inputting bioassay protocol and architecture specification files.

The bioassay protocol incorporates data on how the bioassay is performed; which is in the form of a directed acyclic graph (DAG) of various operations involved in the protocol. The architecture specification file includes various data concerning the size of array of electrodes, location of I/O reservoirs on the periphery of the chip, location specific-purpose modules (e.g. detection, heating) within the array of electrodes and etc. Given the protocol and architecture specification files the synthesis flow is started and passes through various stages, Scheduling, Placement, and Droplet Routing. Figure 3 depicts synthesis flow of a typical DMFB.

Initially, given the information of bioassay protocol and architecture specification files the scheduler algorithm attempts to assign exact start and end times to every operation within the bioassay protocol. This is directly proportional to the size of array of electrodes and the number of available resources to be allocated towards scheduling process.

The scheduler is responsible for making the best use of available resources thus producing the shortest schedule for the bioassay. While attempting to schedule as many operations as possible during any single time-step the scheduler must ensure that the dependencies between operations are met.

The placer algorithm attempts to place scheduled operation onto the array of electrodes given the availability of resources. The placer must attempt to place as many scheduled operations as possible so that all operations scheduled during any given time-step are successfully placed. Given successful placement of operations next the droplet routing stage accounts for routing of droplets onto the surface of the array of electrodes. The stage incorporates routing of droplets from input reservoirs on the periphery of the chip to modules, between the modules and from modules to output reservoirs on the periphery of the chip.

## 2.4 Chip Level Design

Given scheduling, placement and droplet routing stages the pin-mapping and wire routing stages are optional stages towards reduced manufacturing costs. Early generation of DMFBs devoted a dedicated pin per single electrode namely called direct-addressing scheme; in which an array of size m × n requires m + n control pins. The direct-addressing scheme incurres higher manufacturing costs compared with other approaches.

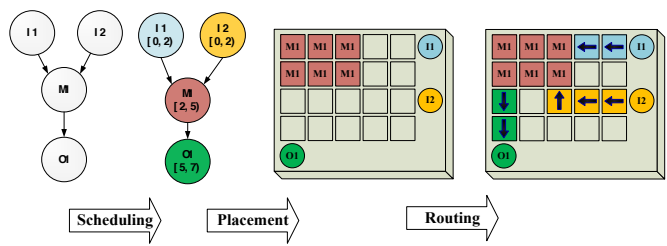

Scheduling　　Placement　　Routing

**Figure 3 Synthesis flow of a typical digital microfluidic biochip (I denotes input, O denotes output reservoirs, and M represents mixing operations)**

Another alternative is cross-referencing scheme which assigns a dedicated control pin per distinct row and column; thus, an array of size m × n would require m + n control pins. Although requiring significantly lower number of control pins than the direct-addressing scheme yet the approaches requires careful control over electrode actuations as concurrent activation of several rows and columns might cause unintended droplet movement.

A more promising approach is active-matrix scheme which requires m + n control pins for addressing an array of size m × n; the approach eliminates the challenge of unintended droplet movement associated with the cross-referencing scheme while at the same time providing high degree of flexibility in terms of droplet movement comparable with direct-addressing scheme. Yet, given the significantly lower number of control pins the manufacturing costs are remarkably reduced.

A handier approach is pin-constrained scheme; as the name implies the scheme constrains the number of control pins thus considerably reducing the manufacturing costs compared with the direct-addressing scheme. The pin-constrained scheme is achieved by initially determining electrodes with equal functionalities; then the electrodes are grouped together, and a single control pin is devoted to the group of electrodes. Though, the manufacturing costs are reduced considerably the scheme suffers lower flexibility in terms of concurrent movement of droplets.

According to the chip architecture, there are two types of placement algorithms; namely, free and fixed placement algorithms. In case of free placement algorithms the placer is responsible for searching for free spaces on the array of electrodes and then assigning the operation with enough free space; whereas in case of fixed placement algorithms the location of modules (where operations are performed) is already fixed and the placer solely binds the operations to the first available location (module), given the type of resource required by the operation.

Given the type of placement algorithm, free or fixed placement, the routing stage would be different. In case of free placement, it is might happen that modules are placed such that there are no routing paths for routing droplets into and out of some modules. In such a case the droplet routing stage fails and as a solution the placement stage must be modified such that possible deadlocks and blockages are eliminated; then, the droplet routing stage is attempted again. In case of fixed placement algorithm as the location of modules is fixed apriori there are dedicated droplet routing paths and subsequently there would be no deadlocks and blockages

Wire routing stage assigns routes to conduction wires between the control pins and the signal pads with a total minimum wire length.

## 3. RELATED WORK

This section briefly reviews prior works on various DMFB designs and architectures; there have been various assay-specific/multi-function DMFB designs yet for the sake of this study only field-programmable general-purpose DMFB designs will be reviewed.

Grissom et al. [8] proposed their field-programmable pin-constrained (FPPC) design towards general-purpose bioassay execution. Also, Grissom et al. [10] proposed an enhanced version of their field-programmable pin-constrained design, namely EFPPC, aimed at reduced manufacturing costs.

Keszocze et al. [9] proposed their DMFB synthesis flow based on general and exact routing methodology; producing DMFB designs with significantly low number of control pins. Yet, their proposed method computationally intensive and feasible in case of small and medium size bioassys.

Wille et al. [11] one-pass synthesis scheme; resulting in much faster computation times compared with Keszocze et al. [9] which in turn allows for performing large bioassays such as protein bioassay. Yet, produced result by their synthesis flow is approximate and highly variable among different runs.

Abdoli et al. [12][13] proposed their general-purpose file-programmable pin-constrained (GFPC) design and its fault-tolerant variant primarily aimed at improved overall bioassay completion times and higher degree of fault-tolerance [14]. They presented their improved version of general-purpose field-programmable pin-constrained (EGFPC) design aimed at reducing overall manufacturing costs while retaining competitive performance with prior designs [15].

Also, Abdoli et al. [16] proposed cell-based field-programmable pin-constrained designs inspired by field-programmable logic array (FPGA) devices.

## 4. THE PROPOSED DESIGN

Figure 4 illustrates the proposed DMFB design; as can be seen, the design incorporates three droplet routing paths along with 4 mixing modules and 8 storage/split/detection (SSD) modules.

As illustrated in Figure 3 the proposed DMFB design consists of three droplet routing paths; routing paths utilize 3-phase routing \cite{srinivasan04}, which simply means only three pins are used for a distinct routing path. Subsequently, pins 1-9 are devoted to vertical left, horizontal and vertical right droplet routing paths.

The proposed DMFB is designed such that SSD modules are integrated into mixing modules; every mixing module accommodates 2 SSD modules at the top left and right corners of the mixing module. Also, mixing modules above and below the horizontal droplet routing path require separate groups of control pins; which is towards reduced overall manufacturing costs.

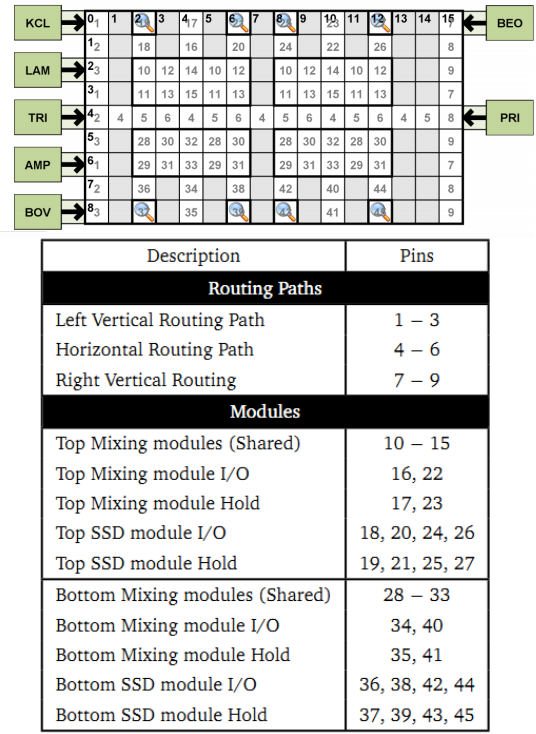

| Description | Pins |
|---|---|
| **Routing Paths** | |
| Left Vertical Routing Path | 1 – 3 |
| Horizontal Routing Path | 4 – 6 |
| Right Vertical Routing | 7 – 9 |
| **Modules** | |
| Top Mixing modules (Shared) | 10 – 15 |
| Top Mixing module I/O | 16, 22 |
| Top Mixing module Hold | 17, 23 |
| Top SSD module I/O | 18, 20, 24, 26 |
| Top SSD module Hold | 19, 21, 25, 27 |
| Bottom Mixing modules (Shared) | 28 – 33 |
| Bottom Mixing module I/O | 34, 40 |
| Bottom Mixing module Hold | 35, 41 |
| Bottom SSD module I/O | 36, 38, 42, 44 |
| Bottom SSD module Hold | 37, 39, 43, 45 |

**Figure 4. The proposed low-cost performance-efficient pin-constrained (LFPC) DMFB design**

The I/O reservoirs are located on the left and right side of the array of electrodes. During the process of moving input droplets from I/O reservoirs to modules initially, droplets are dispensed from left/right I/O reservoirs and moved along the left/right vertical routing paths towards the horizontal routing path; then droplets are moved along the horizontal routing path till reaching below the I/O pin of the intended mixing/SDD module. Then, the droplet is moved up towards the I/O pin on the intended module; subsequently, the droplet is moved onto the I/O electrode. Eventually the droplet is moved to the hold electrode.

Similarly, the same process applies to movement of droplets between mixing/SSD modules; in which initially, the droplet is moved out of the source module through the I/O pin and to the horizontal routing path. Then the droplet is moved along the horizontal routing path till reaching I/O pin of the destination module. Finally, the droplet is moved towards the I/O electrode and then to the hold electrode. The process of moving droplets from mixing/SSD modules to output reservoirs is reverse of the input process; firstly, the droplet is moved out of the module to the horizontal routing path.

Given the location of output reservoirs on the left or right periphery of the array of electrodes the droplet is moved to appropriate left/right vertical routing path. Eventually, droplet is moved along the vertical routing path to reach the intended I/O reservoir.

## 5. Hardware Analysis

This section introduces parameters affecting the overall manufacturing cost of a typical DMFB; the primary focus of this section is to provide detailed cost analysis on wire routing costs of various field-programmable pin-constrained DMFB designs. The key parameters of DMFB wire routing costs include:

- Dimensions of the array of electrodes (Electrode size)
- Number of PCB metal-layers
- Pin-count

Early generation of field-programmable pin-constrained DMFB designs solely focused on general-purpose bioassay execution along with pin-counts as low as possible. Lower pin-count would eliminate the need for additional circuitry for driving control pins; yet sometimes it meant more complex wire routing and higher number of PCB metal layers. Figure 4 depicts wire routing of the proposed low-cost field-programmable pin-constrained DMFB design; as can be seen, the design solely requires one single PCB metal-layer which can greatly reduce the overall wire routing costs.

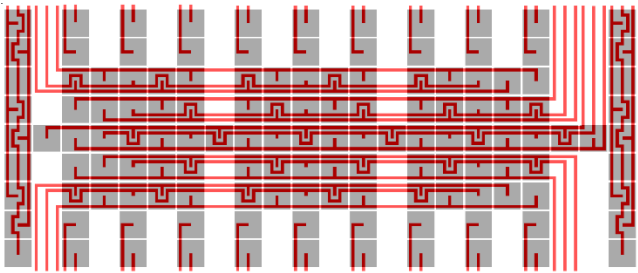

**Figure 4. Wire routing of the proposed low-cost field-programmable pin-constrained DMFB design**

Figure 5 shows the wire routing of enhanced GFPC design (EGFPC) which applies various improvements over the original GFPC design to reduce the number of metal-layers towards wire routing; among which are updated pin-mapping for shared pins inside mixing modules, omission of vertical droplet routing paths in between the mixing modules.

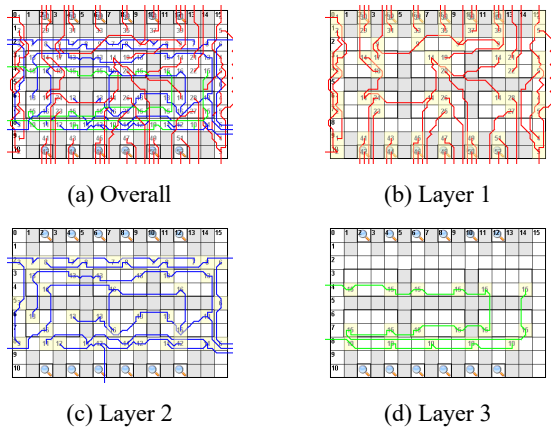

(a) Overall      (b) Layer 1

(c) Layer 2      (d) Layer 3

Figure 5. Wire-routing of the enhanced GFPC architecture with orthogonal capacity of 2.

## 6. CONCLUSION

The wire-routing cost of a DMFB is directly affected by the cost of PCB metal layers plus the cost of additional circuitry; using larger feature sizes tends to reduce the PCB cost in terms of the number of PCB metal-layers.